\documentclass[twocolumn, showpacs, preprintnumbers, amsmath, amssymb] {revtex4}
\usepackage{graphicx}
\usepackage{dcolumn}
\usepackage{bm}

\newcommand{\be}{\begin{equation}}
\newcommand{\ee}{\end{equation}}

\begin{document}

\title{Shear thickening and jamming in densely packed suspensions of different particle shapes}

\author{Eric Brown$^{1,4}$, Hanjun Zhang$^{2,5}$, Nicole A. Forman$^{2,3}$, Benjamin W. Maynor $^3$,\\
Douglas E. Betts$^2$, Joseph M. DeSimone$^{2,3}$, Heinrich M. Jaeger$^1$}

\affiliation{$^1$James Franck Institute, The University of Chicago, Chicago, IL 60637\\
$^2$Department of Chemistry, University of North Carolina, Chapel Hill, NC 27599\\
$^3$Liquidia Technologies, Research Triangle Park, NC 27709\\
$^4$current address: School of Natural Sciences, University of California, Merced, CA 95343\\
$^5$current address: Duracell Technology Center, The Procter \& Gamble Company, Bethel, CT 06801}

\date{\today}

\begin{abstract}

We investigated the effects of particle shape on shear thickening in densely packed suspensions.  Rods of different aspect ratios and non-convex hooked rods were fabricated.  Viscosity curves and normal stresses were measured using a rheometer for a wide range of packing fractions for each shape. Suspensions of each shape exhibit qualitatively similar Discontinuous Shear Thickening.   The logarithmic slope of the stress/shear-rate relation increases dramatically with packing fraction and diverges at a critical packing fraction $\phi_c$ which depends on particle shape.  The packing fraction dependence of the viscosity curves for different convex shapes can be collapsed when the packing fraction is normalized by $\phi_c$.  Intriguingly, viscosity curves for non-convex particles do not collapse on the same set as convex particles, showing strong shear thickening over a wider range of packing fraction.  The value of $\phi_c$ is found to coincide with the onset of a yield stress at the jamming transition, suggesting the jamming transition also controls shear thickening.  The yield stress is found to correspond with trapped air in the suspensions, and the scale of the stress can be attributed to interfacial tension forces which dramatically increase above $\phi_c$ due to the geometric constraints of jamming.  Using this connection we show that the jamming transition can be identified by simply looking at the surface of suspensions.  The relationship between shear and normal stresses is found to be linear in both the shear thickening and jammed regimes, indicating that the shear stresses come from friction.  In the limit of zero shear rate, normal stresses pull the rheometer plates together due to the surface tension of the liquid below $\phi_c$, but push the rheometer plates apart due to jamming above $\phi_c$.

\end{abstract}


 \maketitle

\section{Introduction}

Shear thickening is a category of non-Newtonian fluid behavior in which the viscosity increases as a function of shear rate.  A particularly dramatic form, known as Discontinuous Shear Thickening, occurs in many densely packed suspensions and colloids, the most well-known example being cornstarch in water.  These suspensions feel like a thin liquid when stirred weakly, but feel very thick when stirred harder, and feel thin again when the stress is removed.  Discontinuous Shear Thickening is quantitatively characterized by a steep jump in the shear stress $\tau$ with increasing shear rate $\dot\gamma$, which becomes steeper with increasing packing fraction up to the point where it appears to be discontinuous  \cite{MW58, Ho72, Ba89, MW01a, BJ10}.  The stress jump can be attributed to frictional dissipation from chains of solid particle contacts that span the system due to boundary confinement in response to dilation under shear \cite{BJ10}.  Here we investigate how Discontinuous Shear Thickening depends on the shape of particles in suspension.  Designer particle suspensions are of potential practical interest due to the damping and shock absorbing abilities of Discontinuous Shear Thickening suspensions \cite{LWW03}.

Different particle shapes also provide an opportunity to investigate the relation between jamming and Discontinuous Shear Thickening, since different particle shapes can pack at very different volume fractions $\phi$ \cite{EW05, Philipse96}.  While it has long been suggested that shear thickening is related to jamming  \cite{CWBC98}, there have been few quantitative connections.  In a previous letter, we showed that the packing fraction dependence of the slope of $\tau(\dot\gamma)$ can be characterized as a power law diverging at a critical packing fraction $\phi_c$ for nearly-spherical particles \cite{BJ09}.  Above $\phi_c$ these systems are jammed, meaning they will not flow for applied stresses below a yield stress because geometric constraints from particle contacts suppress the relative shearing of particles \cite{LN98}.  In this paper we extend our work in Ref.~\cite{BJ09} to further characterize the relationship between the mechanics of shear thickening and jammed systems using different particle shapes.


A few previous works have investigated effects of particle shape on shear thickening.  Clarke \cite{Clarke67} compared shear thickening for different shapes, but this was done at a single packing fraction and so did not relate shear thickening to the packing behavior of the particles. Beazley \cite{Beazley80} showed that higher aspect ratio particles shear thickened at lower packing fractions, and suggested that this was related to the lower random packing densities of the higher aspect ratio particles.  More recently, Egres \& Wagner \cite{EW05} performed detailed measurements of shear thickening for colloidal rods of different aspect ratio.  They compared the packing fraction range of Discontinuous Shear Thickening with dense packings obtained from centrifugation, and observed a similar {\em qualitative} trend of decreasing packing fraction with aspect ratio.  Here we go beyond those works in two important ways.  First, using the unique PRINT$^{\textregistered}$ process \cite{RMEEDD05}, we were able to fabricate out of identical materials a wider variety of particle shapes including rods with different aspect ratios, as well as some non-convex rods with hooks.  Second, we show a {\em quantitative} match between critical points for shear thickening and jamming by quantifying the slopes of the shear thickening regime, the yield stress, and normal stresses as a function of packing fraction for different shapes.

The rest of this paper is organized as follows. Rheological measurement methods and particle details are described in Sec.~\ref{sec:methods}.   Viscosity curves for particles of different shapes over a range of packing fractions are shown in Sec.~\ref{sec:visccurves}. We then compare the strength of shear thickening for different shapes based on the logarithmic slopes of $\tau(\dot\gamma)$ as a function of packing fraction $\phi$ for different particle shapes in Sec.~\ref{sec:slopes}, where we show that the packing fraction dependence for many convex shapes can be collapsed relative to the shape-dependent critical packing fraction $\phi_c$.   In Sec.~\ref{sec:trappedair} we show that yield stresses which are mainly measured above $\phi_c$ are correlated strongly with trapped air in the suspensions suggesting the stresses come from interfacial tension.  In Sec.~\ref{sec:surfaceimages} we use images of the surface of suspensions at different packing fractions to further demonstrate the connection between surface tension and the yield stress.  In Sec.~\ref{sec:normalstress} we  show measurements of normal stress in relation to shear stress suggesting a frictional relation, and how the static normal stress changes abruptly at $\phi_c$.  In Sec.~\ref{sec:discussion} we  use the measurements of slopes of the viscosity curves for different packing fractions and shapes to reinterpret previous results that were attributed to particle shape or polydispersity.

\section{Materials and Methods}
\label{sec:methods}

\subsection{Materials}

\begin{figure*}                                                
\centerline{\includegraphics[width=5.5in]{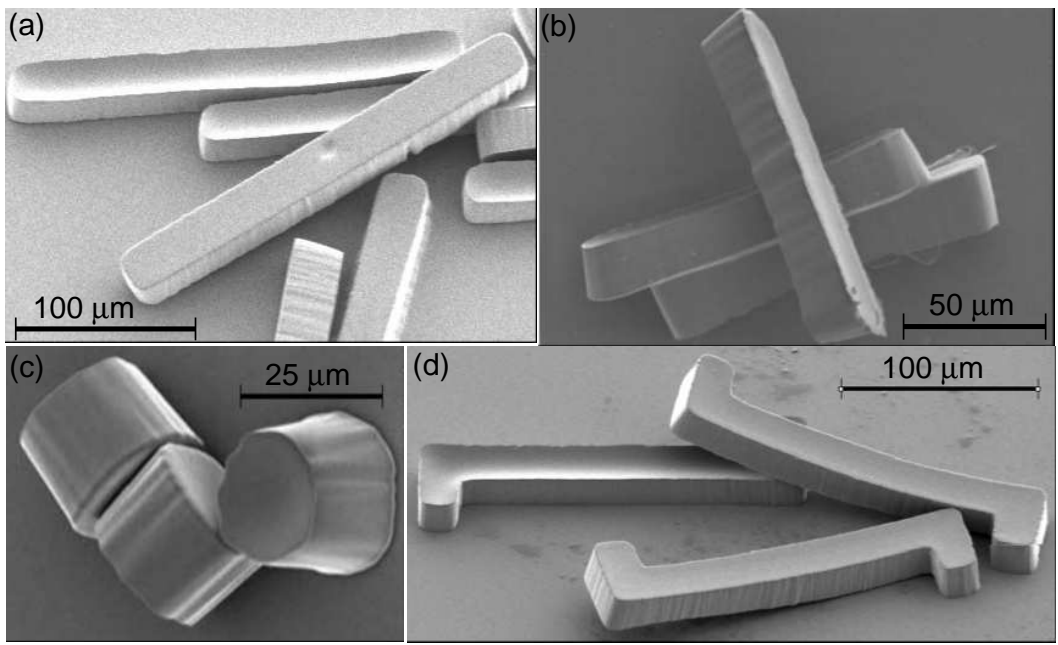}}
\caption{Scanning electron microscope images of dry PEG particles.  (a) Aspect ratio $\Gamma=9$ rods. (b) $\Gamma=6$ rods.  (c) $\Gamma=1$ rods.  (d) $\Gamma=9$ hooked rods}                                        
\label{fig:rodimages}
\end{figure*}

For this study we fabricated particles on the scale of 20-200 $\mu$m to ensure geometric effects were prominent over Brownian motion and interparticle interactions.  We also designed the chemistry specifically to minimize particle-fluid surface tension so the suspension would exhibit shear thickening \cite{BFOZMBDJ10}.  The particles were fabricated using the PRINT$^{\textregistered}$ process \cite{RMEEDD05}, in which molds are made to design the shape of particles.  The particles were composed of 57\% (w/w) triacrylate, 40\% (w/w) methyl ether acrylate, 1\% (w/w) 1-hydroxycyclohexyl phenyl ketone, and 2\% (w/w) fluorescein o-acrylate.  Details of the process used to make the particles for this study can be found in Ref.~\cite{BZFMBDJ10}.  Three types of rectangular rods were used, shown in Fig.~\ref{fig:rodimages}.  The first has dimensions ($266.5 \pm0.6$ $\mu$m) by $ (29.9\pm0.5$ $\mu$m) by $ (31.4\pm0.4$ $\mu$m) referred to as aspect ratio $\Gamma=9$ rods, the second has dimensions ($139.5\pm0.7$ $\mu$m) by $(24.9\pm0.6$ $\mu$m) by $(30.2\pm0.4$ $\mu$m) referred to as aspect ratio $\Gamma=6$ rods, and the third has dimensions ($21.3\pm2.5$ $\mu$m) by $(25.3\pm0.8$ $\mu$m) by $(25.3\pm0.8$ $\mu$m) referred to as aspect ratio $\Gamma=1$ rods.  A fourth shape was fabricated, shown in Fig.~\ref{fig:rodimages}d, which is a variation on the $\Gamma=9$ rods with cubic ``hooks" on opposite ends, so the particle has the shape of an `S'.  These are referred to as hooked rods.  These rods have dimensions ($198.1\pm1.1$ $\mu$m)  by $ (20.6\pm0.4$ $\mu$m)  by $  (24.5\pm0.8$ $\mu$m), and the cubic hooks are $23.0\pm1.2$ $\mu$m long, share the width ($20.6\pm0.4$ $\mu$m) with the rod, and are $24.7\pm1.2$ $\mu$m wide along the length of the rod.

  
The particles were suspended in poly(ethylene glycol) dimethyl ether (PEG) (Mn = 250g/mol) with a viscosity of $\eta_l = 60$ mPa s.  The particles are nearly density matched with the PEG solvent; based on settling times we estimate the density of the particles to be 1\% greater than that of the fluid \cite{BZFMBDJ10}. 


For a comparison to standard spherical particles, and for some direct visualizations, we used opaque polyethylene spheres obtained from Cospheric.  These particles have a nominal diameter range of 125-150 $\mu$m and density of 1.01 g/mL. They were dispersed in silicone oil AR 20 with a nominal density of 1.01 g/mL and viscosity of 20 mPa s.  This suspension is then density matched $\pm 0.01$ g/mL.

\subsection{Methods}

Viscosity measurements were made with an Anton Paar Physica MCR 301 rheometer.  The sample of thickness $d$ is held in a thin layer between two horizontal circular smooth plates of radius $R=12.5$ mm.  The sides of the sample are held at the edge of the plates by surface tension.  The rheometer applies a constant torque $T$ which shears a sample while recording the top plate rotation rate $\omega$.  The viscosity is measured as $\eta\equiv\tau/\dot\gamma$ in a steady state for shear stress 

\be
\tau = \frac{2T}{\pi R^3}
\label{eqn:stress}
\ee

\noindent and shear rate 

\be
\dot\gamma = \frac{R\omega}{d} \ .
\label{eqn:shearrate}
\ee

\noindent  The normal stress $\tau_N$ is measured as the normal force pushing against the top plate divided by the cross-sectional area of the plate. These definitions are meant to characterize the average mechanical response due to an external force in a size-independent way.  In systems that exhibit Discontinuous Shear Thickening, this definition can differ dramatically from the local hydrodynamic constitutive relation between shear stress and shear rate because the shear stress is strongly coupled to normal stresses which depend on boundary conditions \cite{BJ10}.  


Measurements were made at a bottom plate temperature controlled at $20^{\circ}$ C with the room humidity ranging from 22\% to 38\%, although during individual experiments the humidity was constant.  This affects solvent evaporation/adsorption which can have a significant effect on the rheology due to the sensitive packing fraction dependence of shear thickening suspensions \cite{BJ09}.  To minimize evaporation or adsorption, we used a solvent trap which enclosed the sample and a small amount of air around it by an extra layer of liquid.  

Packing fractions $\phi$ were defined as the volume of solid particles over the total volume of solids plus liquid.  Each ingredient was measured by mass which was divided by the material density to obtain the volume.  The particles and liquids were stirred together for 30 s using a vortex mixer.  The packing fraction in terms of the inverse of the available free volume per particle decreases slightly during measurements as the samples are seen to dilate during shear thickening \cite{BJ10}.  There may also be a difference in packing fraction definitions due to air bubbles becoming trapped in the interior of the suspension (see Sec.~\ref{sec:trappedair}).  Because of the small size of our samples and changes over time due to humidity, our absolute measurements of packing fraction could vary by as much as 0.04 for each shape upon trying to reproduce mixtures.  This is the relevant error for comparing to external experiments.  Within an individual measurement series where the packing fraction was lowered by adding a small amount of liquid PEG for each successive experiment, the uncertainty in packing fraction when comparing to neighboring data points within that series was estimated to be 0.005.


Samples were pre-sheared immediately before measurements for at least 100 seconds at shear rates above the shear thickening regime where the steady state flow is fully mobilized \cite{BJ10}.  After this pre-shear, measurements on suspensions were found to be repeatable with a typical variation of 10-20\% from run to run.   This is the variation whether we remeasure a sample that is in place on the rheometer or replace it with a new one with the same procedure.   Following the preshear, measurements were performed with slowly decreasing and then increasing stress ramps and additional runs were made with different control ramp rates to check for hysteresis, thixotropy, and transients.  Examples of hysteresis loops for different ramp rates are shown in Ref.~\cite{BJ10}.  In this paper, viscosity curves are shown for a ramp rate of 500 s per decade of stress which is slow enough to be in the steady state limit.  We show only one set of curves for brevity since in the steady state limit they were all identical within typical variations.

Because only small samples of particles could be fabricated, the sample thickness $d$ was limited to no more than a few particles across.  We ensured that in each experiment, $d$ was more than twice the longest length the particle, which is large enough to avoid significant finite-size effects on the viscosity curves \cite{BZFMBDJ10}.  Slip was shown not to have a significant effect on viscosity curves for similar measurements with some of the same particles \cite{BZFMBDJ10}.  The suspensions were observed not to spill out of the rheometer until shear rates exceeding those of the reported measurements; however in most cases this set the upper limit of the range of measurements.

\section{Viscosity curves}
\label{sec:visccurves}

\begin{figure}                                                
\centerline{\includegraphics[width=3.4in]{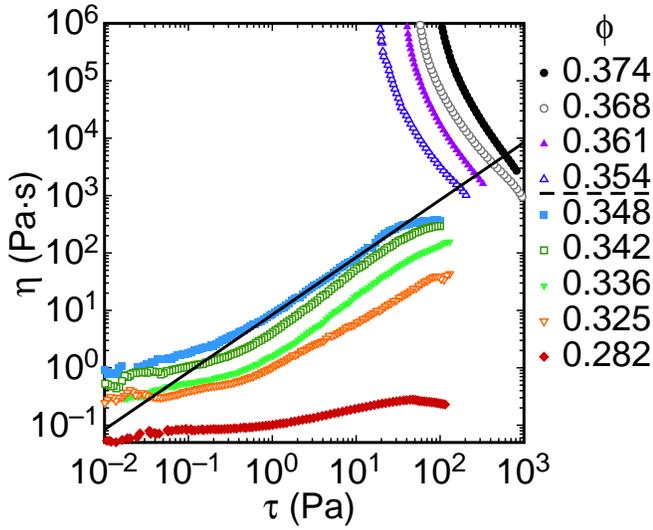}}
\caption{(color online) Viscosity vs.~shear stress for rods of aspect ratio $\Gamma=9$.  Packing fractions $\phi$ are decreasing from top to bottom. The solid line has a slope of 1 corresponding to a constant shear rate.  The dashed line in the key indicates the critical packing fraction $\phi_c$ above which the suspension is jammed with a large yield stress, and below which the suspension exhibits shear thickening.}
\label{fig:viscstress_Gamma12}                                        
\end{figure}

\begin{figure}                                                
\centerline{\includegraphics[width=3.4in]{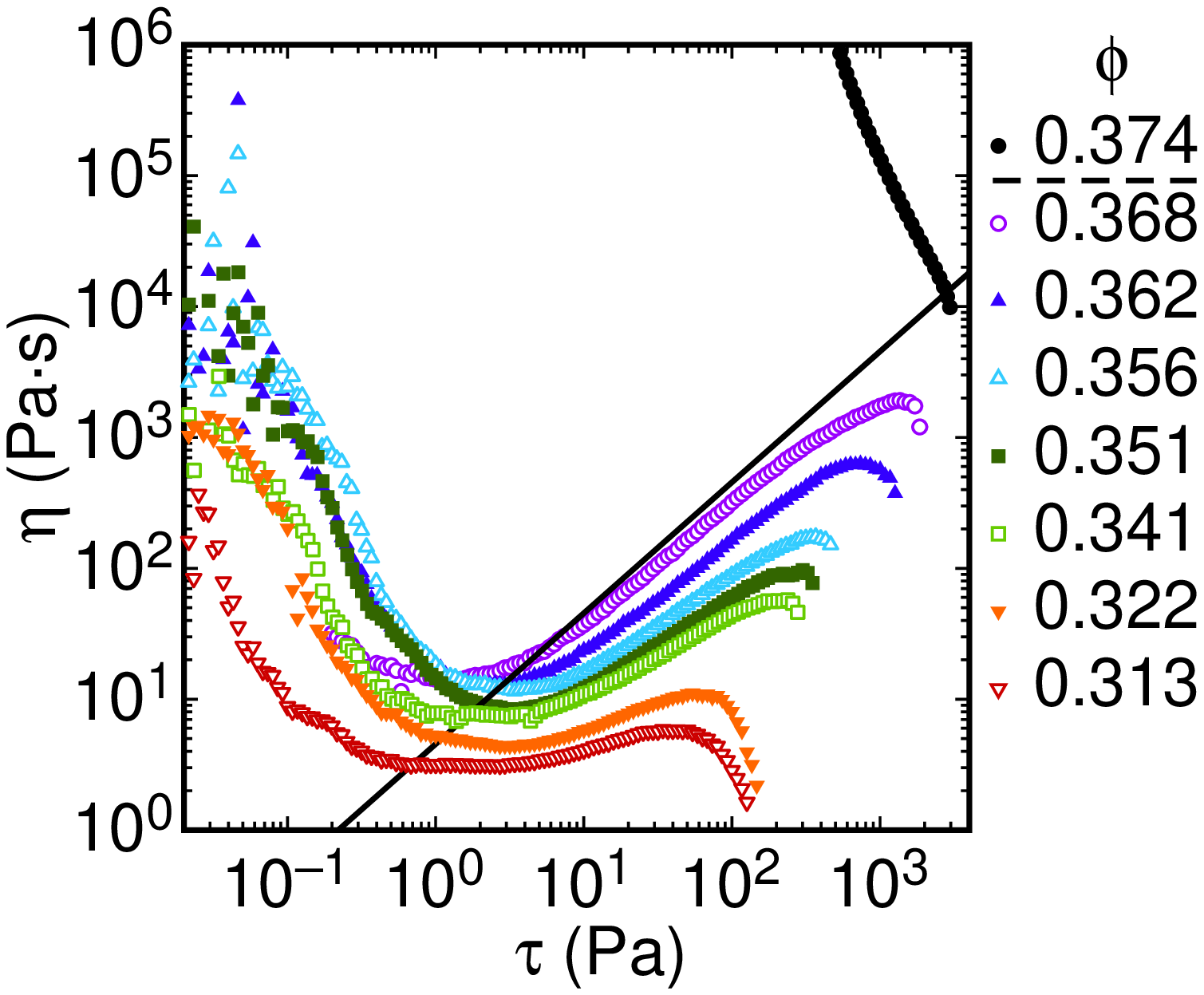}}
\caption{(color online) Viscosity vs.~shear stress for rods of aspect ratio $\Gamma=6$.  Packing fractions $\phi$ are decreasing from top to bottom. The solid line has a slope of 1 corresponding to a constant shear rate.  The dashed line in the key indicates the critical packing fraction $\phi_c$.
}
\label{fig:viscstress_Gamma6}                                        
\end{figure}

\begin{figure}                                                
\centerline{\includegraphics[width=3.4in]{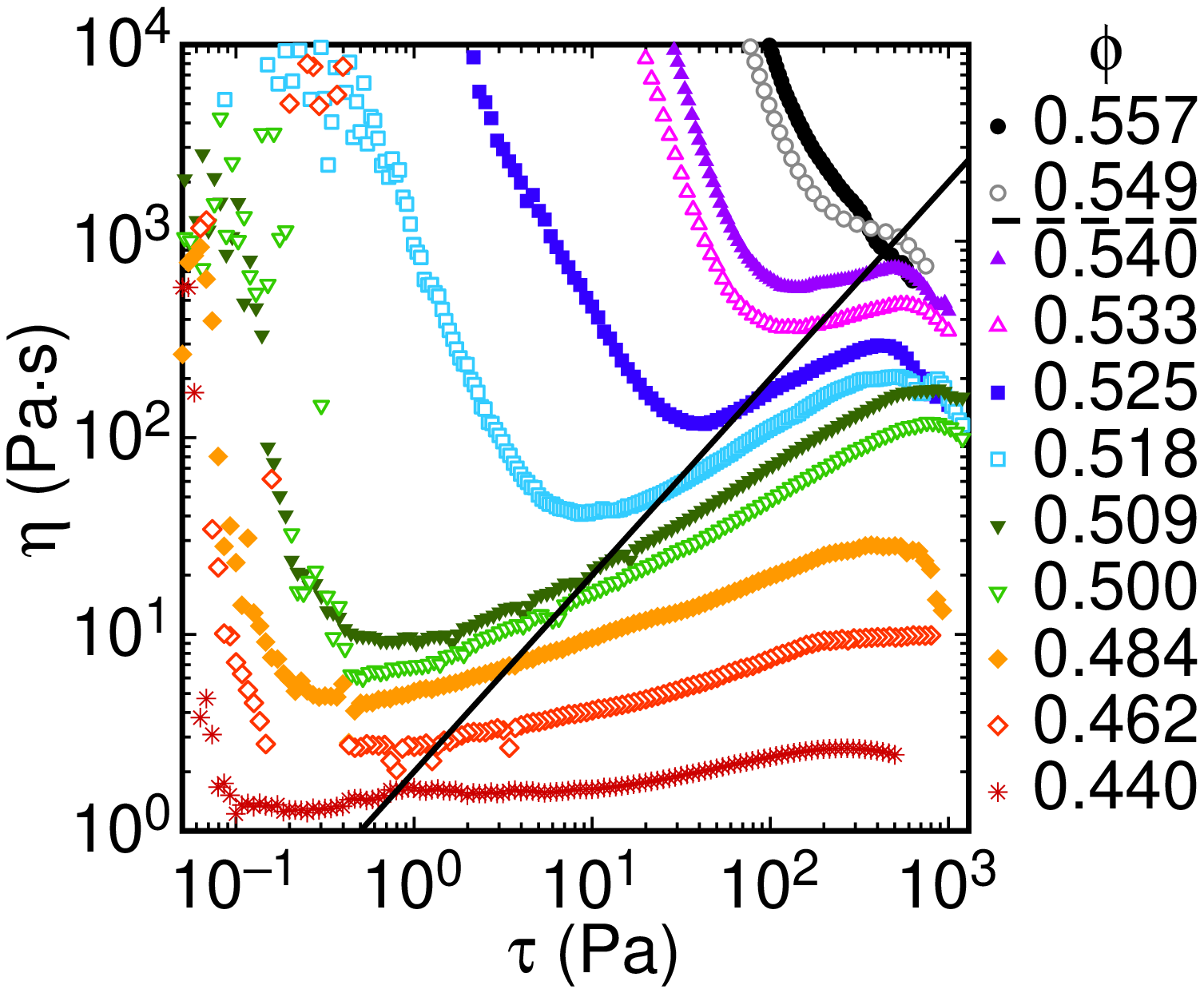}}
\caption{(color online) Viscosity vs.~shear stress for rods of aspect ratio $\Gamma=1$.  Packing fractions $\phi$ are decreasing from top to bottom. The solid line has a slope of 1 corresponding to a constant shear rate.  The dashed line in the key indicates the critical packing fraction $\phi_c$.}
\label{fig:viscstress_Gamma1}                                        
\end{figure}

\begin{figure}                                                
\centerline{\includegraphics[width=3.4in]{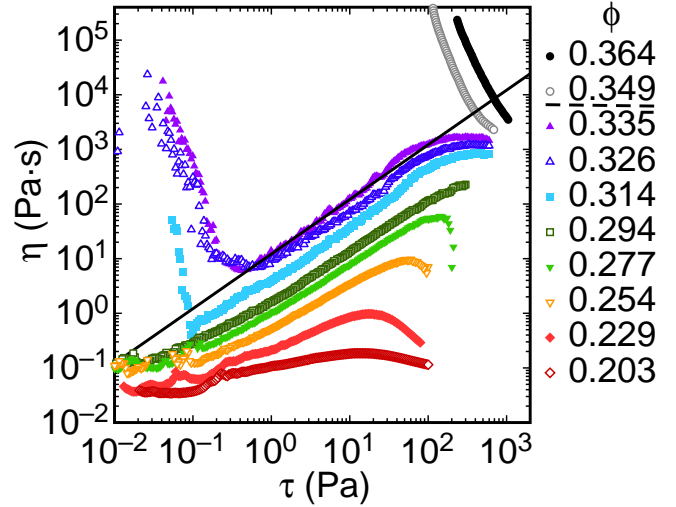}}
\caption{(color online) Viscosity vs.~shear stress for $\Gamma=9$ hooked rods.  Packing fractions $\phi$ are decreasing from top to bottom. The solid line has a slope of 1 corresponding to a constant shear rate.  The dashed line in the key indicates the critical packing fraction $\phi_c$.}
\label{fig:viscstress_hookedrods}                                        
\end{figure}

In this section we show viscosity curves as viscosity vs. shear stress for suspensions of particles with different shapes over a range of packing fractions.  Viscosity curves for rods of aspect ratios $\Gamma=9$, $\Gamma=6$, $\Gamma=1$, and $\Gamma=9$ hooked rods are shown in Figs.~\ref{fig:viscstress_Gamma12}, \ref{fig:viscstress_Gamma6},  \ref{fig:viscstress_Gamma1}, and \ref{fig:viscstress_hookedrods} respectively, for different packing fractions $\phi$.  In each case, the viscosity curves exhibit a shear thickening regime indicated by a positive slope in some intermediate stress range which does not vary too much with packing fraction.  The slope of this shear thickening regime becomes steeper with increasing packing fraction, ultimately reaching a maximum slope of about 1 on a log-log plot of viscosity vs.~stress.  Because of the definition $\eta=\tau/\dot\gamma$ this limit corresponds to a constant shear rate or equivalently an infinite slope of shear stress vs.~shear rate.  At higher packing fractions, instead of shear thickening, a large yield stress is observed, which is indicated by the viscosity diverging as the shear stress decreases. This behavior is typical of Discontinuous Shear Thickening in nearly-spherical particles \cite{MW01a, BJ09, BJ10} or rods \cite{EW05}.

One notable feature of the viscosity curves of Figs.~\ref{fig:viscstress_Gamma12}, \ref{fig:viscstress_Gamma6}, \ref{fig:viscstress_Gamma1}, and \ref{fig:viscstress_hookedrods} that is dependent on particle shape is the variation in the packing fraction $\phi_c$ where jamming occurs.  Jamming is usually characterized by the onset of a yield stress in ideal systems \cite{OLLN02}.   Practically, we measure the onset of jamming as the packing fraction where the largest increase in the yield stress occurs \cite{BJ09}, indicated in by the dashed lines in the keys of Figs.~\ref{fig:viscstress_Gamma12}, \ref{fig:viscstress_Gamma6},  \ref{fig:viscstress_Gamma1}, and \ref{fig:viscstress_hookedrods}.  The very small yield stresses at lower packing fractions are too weak to attribute to geometric jamming (see Sec.~\ref{sec:yieldstress} for a detailed investigation).  While suspensions of frictional spheres jam at about $\phi_c=0.58$ \cite{BJ09}, we find $\phi_c=0.55$ for $\Gamma=1$ rods, $\phi_c=0.37$ for $\Gamma=6$ rods, $\phi_c=0.35$ for $\Gamma=9$ rods, and $\phi_c=0.34$ for the hooked rods.  The trend of decreasing packing fraction with increasing rod aspect ratio is consistent with other measurements \cite{EW05, Philipse96}.

\subsection{Packing fraction dependence of slopes}
\label{sec:slopes}


To quantify the strength of shear thickening, we consider the steepness of the viscosity curves following our method in Ref.~\cite{BJ09}.  For each viscosity curve, we fit $\eta\propto\tau^{1-\epsilon}$  (equivalent to $\tau\propto\dot\gamma^{1/\epsilon}$) to the shear thickening regime.  The value of $\epsilon$ characterizes the strength of shear thickening such that the bound  $\epsilon=1$ corresponds to a Newtonian fluid and the bound $\epsilon=0$ corresponds to the steepest possible shear thickening curve with an infinite slope of stress vs.~shear rate.   In Ref.~\cite{BJ09} we measured that $\epsilon$ goes to zero in the limit as $\phi$ approaches $\phi_c$ from below for both suspensions of glass spheres and cornstarch particles.  The critical packing fraction $\phi_c$ was also found to occur at the same packing fraction as the divergence of the magnitude of the viscosity and the onset of the yield stress.  In cases where there is a large yield stress that hides much of the shear thickening regime so the stress-shear rate relation did not exhibit a constant slope over a wide range, we had to modify this procedure to measure the underlying differential increase in stress with shear rate.  In such cases we fit 

\begin{equation}
\tau(\dot\gamma) =  \tau_y+a_1\dot\gamma^{1/2}  + a_2\dot\gamma^{1/\epsilon} 
\label{eqn:HBmodel_plus}
\end{equation}

\noindent to the data in the steep part of the shear thickening regime and at lower stresses \cite{BFOZMBDJ10}.   The first two terms of Eq.~\ref{eqn:HBmodel_plus} characterize the yield stress and shear thinning behavior as in a Herschel-Bulkley model.   When the yield stress and correspondingly the first two  terms of Eq.~\ref{eqn:HBmodel_plus} are small, this procedure is equivalent to the original procedure of fitting just the shear thickening regime to $\tau\propto\dot\gamma^{1/\epsilon}$.

\begin{figure}                                                
\centerline{\includegraphics[width=3.4in]{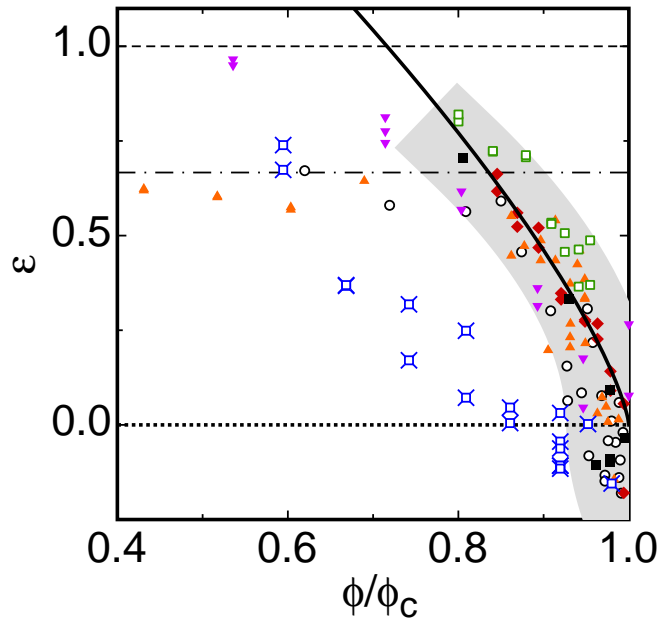}}
\caption{(color online) The strength of shear thickening is characterized by a fit of $\tau\propto \dot\gamma^{1/\epsilon}$ to the shear thickening regime, and $\epsilon$ is plotted vs.~normalized packing fraction $\phi/\phi_c$ for different particle shapes.  Black solid squares:  rods with aspect ratio $\Gamma=9$, $\phi_c=0.35$.  Red diamonds: rods with aspect ratio $\Gamma=6$, $\phi_c=0.37$.   Green open squares: rods with $\Gamma=1$, $\phi_c=0.55$.  Blue  crossed squares:  hooked rods, $\phi_c=0.34$.  Black open circles: glass spheres in water, $\phi_c=0.58$ \cite{BJ09}.  Purple down-pointing triangles: cornstarch in glycerol, $\phi_c=0.57$ \cite{BJ10}.  Orange up-pointing triangles: cornstarch in water, $\phi_c=0.48$ \cite{BJ10}.  The collapse of the data in the gray band suggests that the normalized packing fraction $\phi/\phi_c$ determines the strength of shear thickening for convex particle shapes.  The dashed line at $\epsilon=1$ corresponds to viscous flow with $\tau\propto \dot\gamma$, the limit at small $\phi$ if inertial effects are negligible.  The dashed-dotted is $\epsilon=2/3$, corresponding to an inertial flow with $\tau\propto\dot\gamma^{3/2}$.  The solid line is a best fit of $\epsilon \propto (\phi_c-\phi)^{\xi}$ to the data for convex shapes and $\phi/\phi_c > 0.8$.}
\label{fig:epsilon_phirel}                                        
\end{figure}

Here we apply this same analysis to suspensions of particles of different shapes.   In Fig.~\ref{fig:epsilon_phirel} we plot $\epsilon$ vs. a normalized packing fraction $\phi/\phi_c$.  The value of $\phi_c$ is measured independently for each shape as the packing fraction where there is a steep jump in the yield stress.  Data is shown for each of the shapes studied here, as well as glass spheres from Ref.~\cite{BJ09} and cornstarch from Ref.~\cite{BJ10}.  With the exception of the hooked rods, the data for every convex shape collapse near $\phi_c$, highlighted by the shaded band in Fig.~\ref{fig:epsilon_phirel}. This suggests that the normalized packing fraction $\phi/\phi_c$ generally characterizes shear thickening for each of those convex shapes, despite the large differences in $\phi_c$.  Since the critical packing fraction $\phi_c$ is measured based on the yield stress, which is independent from the slope of the viscosity curve, the collapse confirms a connection between shear thickening and the jamming transition. Specifically, the fact that the slope diverges ($\epsilon \rightarrow 0$) as $\phi$ goes to $\phi_c$ suggests the proximity to the jamming transition controls the strength of shear thickening like a critical point in a second order phase transition.  The scatter from repeated runs is similar to the scatter around zero near $\phi_c$, indicating the typical uncertainty in fitting the viscosity curves to obtain $\epsilon$.  While there has been some success in characterizing the order parameter and scaling exponents for the jamming transition \cite{OT07}, how critical scaling theory applies to Discontinuous Shear Thickening is still an open problem.  The situation is complicated by the fact that these suspensions are neither in equilibrium nor homogeneous, nor does the divergent scaling apply to local constitutive laws \cite{BJ10}.

The collapse of the slopes in Fig.~\ref{fig:epsilon_phirel} suggests that some dense suspensions that show strong shear thickening but do not exhibit a discontinuous stress/shear-rate relation ($\epsilon \rightarrow0$) as in Fig.~\ref{fig:viscstress_Gamma1} may still fit into the same model as Discontinous Shear Thickening suspensions.  It has been shown that the stress contribution from the yield stress can be added linearly to the stress contribution responsible for Discontinuous Shear Thickening \cite{BFOZMBDJ10}.  Since the total viscosity from the sum of these two terms is defined as an absolute ratio $\eta = \tau/\dot\gamma$, the yield stress can make a significant contribution at higher shear rates and the total viscosity can be less strongly shear thickening than the differential increase in stress with shear rate from shear thickening mechanisms.  Based on this interpretation, the large yield stresses near $\phi_c$ in Fig.~\ref{fig:viscstress_Gamma1} hide much of the otherwise shear thickening regime, and the apparent slope of the viscosity curve is lessened by a broad crossover from the dominance of the yield stress to the dominance of the shear thickening term.  This could mean that the stresses from shear thickening mechanisms are the same as in other dense suspensions, but if the stresses from yield stress mechanisms are large they can be dominant in the overall rheology.

To quantify the possible critical scaling behavior, we fit the function $\epsilon \propto (\phi^*-\phi)^{\xi}$ to all of the data for the convex shapes in Fig.~\ref{fig:epsilon_phirel} for $\phi/\phi_c > 0.8$.  Statistical uncertainties of 0.015 on $\phi$ and 0.09 on $\epsilon$ are based on repeated measurements.  Varying the fit value of $\phi^*$, the exponent $\xi$, and the proportionality, we obtain $\phi^*/\phi_c=0.979\pm 0.004$ and $\xi=0.52\pm 0.06$.  Limiting the range of the fit to higher $\phi$, or varying the relative uncertainties on packing fraction and $\epsilon$, does not change the fit values significantly.  Because we can only obtain measurements of $\epsilon$ when $\phi < \phi_c$, only negative errors in $\phi$ are measured near $\phi_c$, which biases the distribution of data near the critical point so the best fit value could be smaller than the actual value by about the uncertainty on $\phi$.  Given this bias, the packing fraction where the viscosity curve becomes discontinuous and $\epsilon$ goes to zero (represented by $\phi^*$) is consistent with the packing fraction where the large yield stress onsets.  If we then redo the fit while fixing $\phi^* = \phi_c$, then we obtain $\xi = 0.74\pm 0.09$.


Further away from $\phi_c$, the critical point no longer dominates the rheology and different forces can be dominant.  For higher viscosity fluids, viscous forces can dominate, where a Newtonian behavior is expected with $\epsilon=1$ (shown by the dashed line in Fig.~\ref{fig:epsilon_phirel}), which is satisfied for example by cornstarch in a glycerol-water mixture at small $\phi$ ($\eta_l = 80$ mPa s, data from Ref.~\cite{BJ10}).  For lower viscosity fluids, inertial forces can dominate in the stress regime of our measurements.  For cornstarch in water ($\eta_l = 1$ mPa s), this results in an intermediate Reynolds number regime in the stress range of interest where $\tau \propto \dot\gamma^{3/2}$, corresponding to $\epsilon=2/3$ (shown by the dashed-dotted line in Fig.~\ref{fig:epsilon_phirel}, data from Ref.~\cite{BJ10}).

\section{Yield stress}
\label{sec:yieldstress}

\begin{figure}                                                
\centerline{\includegraphics[width=3.4in]{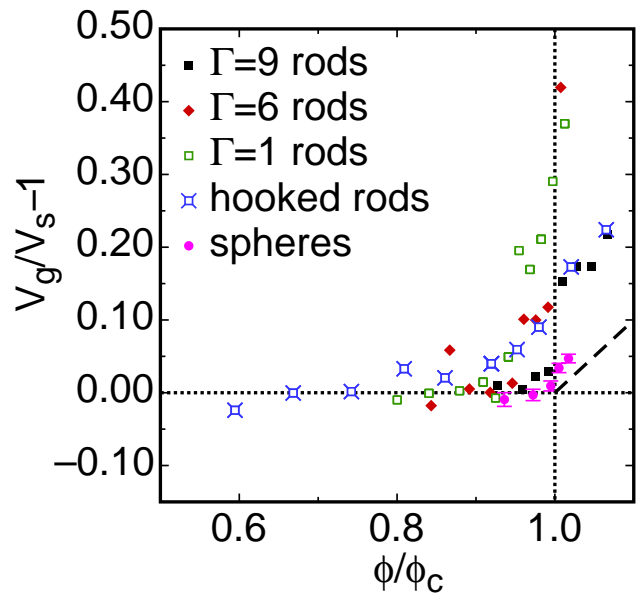}}
\caption{(color online) (a) Scatter plot of yield stress normalized by the surface tension stress scale $\gamma/a$ vs.~normalized packing fraction $\phi/\phi_c$ for PEG particles with different shapes identified in the key, along with data for polyethylene spheres (pink circles).  (b)Normalized difference between the volume of the suspension in the rheometer gap $V_g$ and the volume of solids and liquids mixed in suspension $V_s$, indicating the fractional volume of air trapped in suspension.  Dashed line:  minimum fraction of trapped air due to geometric constraints for hard particles.  The absolute uncertainty in the volume ratio on the vertical scale is 0.02 except for the spheres. 
}
\label{fig:excessphi}                                        
\end{figure}

Here we address the origin of the yield stress seen in the viscosity curves in Figs.~\ref{fig:viscstress_Gamma12}, \ref{fig:viscstress_Gamma6},  \ref{fig:viscstress_Gamma1}, and \ref{fig:viscstress_hookedrods}.  We show a scatter plot of yield stress $\tau_y$ vs.~normalized packing fraction $\phi/\phi_c$ in Fig.~\ref{fig:excessphi}a.  The large yield stress above $\phi_c$ suggests a jammed state, which is attributed to geometric constraints on particle motion when the packing fraction becomes so high that particles cannot move around each other \cite{LN98}.  In systems of confined volume, the scale of the yield stress is traditionally attributed to the stiffness of particles \cite{OLLN02}.  However, for our suspensions, the measured value of the yield stress is only enough to compress hard elastic spheres by a strain of order $\sqrt{\tau_y/E_p} \stackrel{<}{_\sim} 10^{-4}$ using a Hertzian contact model where $E_p$ is the Young's modulus of the particles \cite{OLLN02}.  Under this model, our suspensions could not be compressed to packing fractions above the jamming transition by more than $10^{-4}$.  The smaller yield stresses for $\phi<\phi_c$ in Fig.~\ref{fig:excessphi}a also lacks an explanation.  

\subsection{Trapped air}
\label{sec:trappedair}

The key difference from those models is that our suspensions are not confined to a fixed volume; instead they are confined by surface tension at the liquid-air interface which is much softer than the particles.  If the particles are packed to a density just above the jamming transition, then the liquid-air interface could deform by a distance up to about half a characteristic particle diameter $a$ (calculated as the cube root of the volume) to accommodate the particle packing.  This finite-size effect would only allow the measured packing fraction to reach up to only $a/R \approx 0.004$ above jamming in our case.  These limitations imply that even if we try to mix higher ratios of solids to liquid plus solids, there is not enough confining stress to pack the particles much more densely than at the jamming transition, and so the particle packing must take up more volume by trapping air bubbles within the packing.  We propose that the interfacial tension from these air bubbles and the deformed liquid-air interface at the boundary can be a source of yield stress.  A curved liquid-air interface produces a force from interfacial tension on the attached particles.  If solid frictional contacts between particles span the system, these forces can transmit along chains and be redirected in a random packing such that the resistance to shear stress is proportional to the normal stress due to confinement as in a frictional scaling.  Such force chains are expected at packing fractions above the jamming transition \cite{CWBC98}, at lower packing fractions with even tiny interaction forces between particles \cite{TPCSW01}, and during shear thickening when packings dilate \cite{BJ10}.  The scale of the stress from surface tension scales as $\gamma/r$ for a surface with curvature of radius $r$.  Presuming the scale of the curvature is set by the particle size, this gives a stress scale $\gamma/a$ which we can compare yield stress measurements to.

We measure the volume of trapped air based on the difference between the volume that the suspension takes up in the rheometer gap $V_g$ with an  uncertainty of 2\%, and the volume of solids plus liquids mixed in suspension $V_s$.  We show this difference in Fig.~\ref{fig:excessphi}b as a function of packing fraction for different particle shapes, with the difference in volumes normalized as $V_g/V_s-1$ indicating the fractional volume of air over liquids plus solids in the suspension.  It can be seen that the amount of trapped air is consistent with zero at low packing fractions, but can become somewhat positive as $\phi/\phi_c$ approaches 1 from below and typically jumps to large positive values for $\phi/\phi_c > 1$.  This confirms that there is air trapped in the suspensions above jamming.  The amount is considerably more than the minimum required by geometric constraints on hard particles which will not pack more densely than at the jamming transition (dashed line in Fig.~\ref{fig:excessphi}).  Additionally, it is seen that some air tends to become trapped in the suspensions even below jamming but at high packing fractions where the effective viscosity happens to be large.   The tendency for air to be trapped in suspensions that are jammed or nearly so is likely due to the difficulty of bubbles escaping between particles that are closely packed in suspensions with a high viscosity or yield stress.  We also plot in Fig.~\ref{fig:excessphi} the yield stress and fraction of trapped air for suspensions of 135 $\mu$m polyethylene spheres in mineral oil whose volumes were measured much more precisely.  These samples were sonicated after mixing to increase the mobilization of particles and help trapped air bubbles escape.  These data are seen to match more closely to the minimum amount of trapped air plotted as the dashed line, which suggests the amount of trapped air can vary with the mixing procedure. 

By comparing panels a and b of Fig.~\ref{fig:excessphi}, it can be seen that there is a strong correlation among the different shapes between the yield stress and the amount of trapped air for $\phi<\phi_c$.  In particular, the particle with both the largest yield stress and largest amount of trapped air, the $\Gamma=1$ rods, is the smallest particle used, which might enhance the tendency to trap air bubbles because stresses from surface tension are relatively larger for smaller particles.

\begin{figure}                                                
\centerline{\includegraphics[width=3.4in]{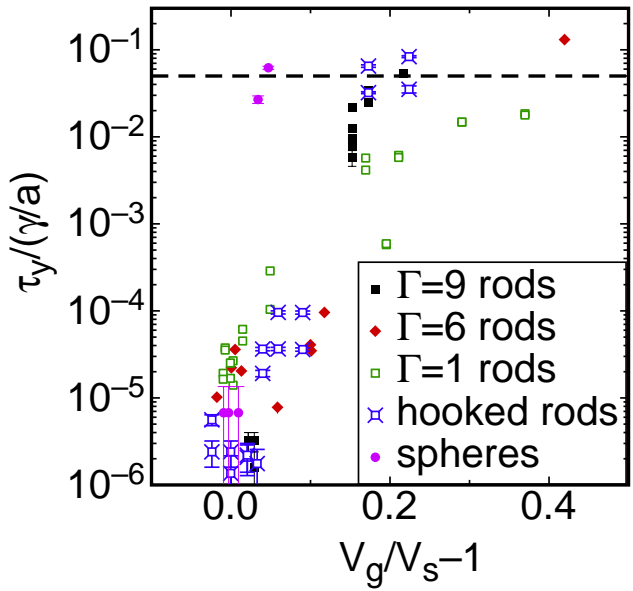}}
\caption{(color online) Scatter plot of yield stress vs.~volume of air trapped in rheometer as a fraction of solid plus liquid sample volume for a variety of PEG particle shapes listed in the key.  Data for polyethylene spheres (pink circles).  For a given series of one particle shape, packing fraction generally increases to the right, and the cluster of points in the lower left corresponds to packing fractions below jamming. Dashed line:  $0.05\gamma/a$ corresponding to the limiting stress value seen in foams jammed by interfacial tension \cite{GDJC98}, 3-phase suspensions \cite{KW11}, or the maximum stress seen in the shear thickening regime \cite{BJ10}.
}
\label{fig:yieldstressair}                                        
\end{figure} 

To connect the trapped air to the yield stress, we show a scatter plot of yield stress $\tau_y$ vs. trapped air fraction $V_g/V_s-1$ in Fig.~\ref{fig:yieldstressair}.  The data correspond to different packing fractions and particle shapes.  The yield stress values are normalized by the predicted interfacial stress scale $\gamma/a$.  A strong positive correlation is seen between the yield stress and volume of trapped air, and the data for different shapes even collapse in a narrow band.  The data clustered at the lower left generally correspond to $\phi <\phi_c$ where the yield stress is small or below the resolution limit and there is little or no measurable trapped air.  The data above $\phi_c$ where the yield stress is large generally have large volumes of trapped air so show up on the upper right of the figure.  The scale of the yield stress in the limit of larger volumes of trapped air is reminiscent of other systems jammed by interfacial tension.  In jammed foams, the yield stress is observed to be on the scale of $\tau_y \approx 0.05\gamma/a$ \cite{GDJC98}.  In 3-phase suspensions with 2 non-miscible fluid phases, a yield stress was observed to be on the scale of $\tau_y \approx 0.05\gamma/a$ for moderate fractions of the 3rd phase \cite{KW11}.  While the 3rd phase in that case was a liquid instead of gas, the forces from interfacial tension are much stronger than gravity on the scale of the particle size.  The maximum stress seen in the shear thickening regime is also found to be on the order of $0.05\gamma/a$ \cite{BJ10}.  While in the latter case the suspensions are not statically jammed, the measured forces are similar because they are transmitted through the suspension along solid contacts between particles, which form when the suspension dilates under shear against the interfacial tension at the boundary.  The similarity with these other results strongly suggests that the value of the yield stress in these jammed suspensions comes from interfacial tension.  



While the specific values of the yield stress and volume of trapped air shown in Fig.~\ref{fig:excessphi} likely depend on the sample preparation procedure, the collapse of yield stress data for the PEG particles in Fig.~\ref{fig:yieldstressair} suggests a general relation between the yield stress and fraction of trapped air, despite the different particle shapes.  If the yield stress is given by the average capillary stress $\gamma/r$ percolated across the suspension along force chains, then the data collapse suggests the average ratio of particle size to bubble size is similar from shape to shape.  While it is not obvious whether the characteristic bubble size is determined by hydrostatics or shear, we note that one of the contributions of a preshear in bringing a sample to a reproducible state could be in bringing the bubble size distribution to a steady state.

\subsection{Surface roughness and capillary forces}
\label{sec:surfaceimages}

If a suspension is jammed, the boundaries must be able to balance the forces that are transmitted along force chains through the bulk.  When the boundary is a liquid air interface, it must  then become curved so that surface tension can provide the stress to push back on particles which penetrate the liquid-air interface \cite{CHH05}.  If the particles are between about 1 and 100 $\mu$m, then the grains are large enough to scatter light diffusively and small enough that they cannot be seen individually, so the surface appears rough by eye.  On the other hand, if a suspension is unjammed there is free space for the particles to rearrange, so any particles on the surface will be pushed by surface tension to the interior (assuming the liquid wets the particles, which is also a requirement to observe shear thickening \citep{BFOZMBDJ10}), resulting in a flat and shiny surface.  In this section we demonstrate this contrast in surface appearance between jammed an unjammed states and that it can be used to make quantitative measurements of the jamming transition $\phi_c$.



\begin{figure*}                                                
\centerline{\includegraphics[width=6.5in]{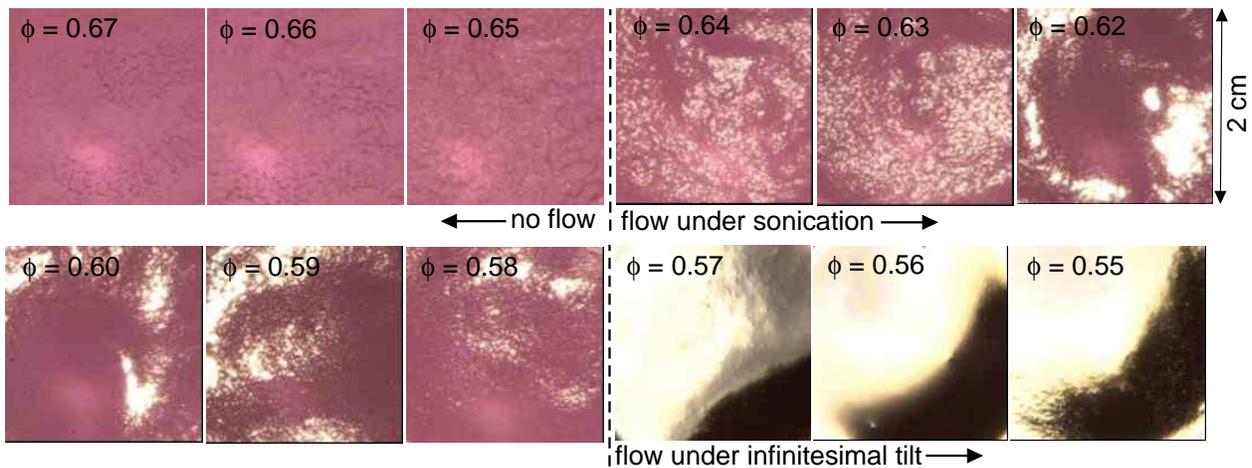}}
\caption{(color online) Top views of a static layer of 135 $\mu$m polyethylene spheres density matched in silicone oil.  The layer was 1.3 mm deep and each image is 2 cm on a side.  Images were taken for different packing fractions given in the corner of each panel.  Lighting was direct to emphasize changes in reflectivity.  Packing fractions $\phi > 0.64$ did not flow under any amount of tilt.  Packing fractions $0.64 \ge \phi > 0.58$ flow while being sonicated, but after stopping sonication, these suspensions no longer flowed at small tilt angles.  Packing fractions $\phi \le 0.57$ flow under even small tilt ($\sim 1$ degree).  The reflectance of the surface increased so much when $\phi$ was decreased to $0.57$ that the camera became oversaturated by the direct reflection of the light.}
\label{fig:surfaceimages}                                        
\end{figure*}

The suspensions used consisted of 135 $\mu$m polyethylene spheres density matched in silicone oil, which are opaque and do not settle.  These suspensions were placed in a  $H=1.3$ mm deep layer and viewed from the top with direct lighting as shown in Fig.~\ref{fig:surfaceimages}.  The series of images was taken by starting at a high packing fraction well above the jamming transition ($\phi> \phi_c$), then adding oil to reduce the packing fraction for each successive image.  At each packing fraction the sample was vibrated at 40kHz with a sonicator for 1 min, tilted during and after sonication to observe whether or not shear occurred, then any flow was allowed to come to rest before the image was taken.  

For the highest packing fractions $\phi > 0.64$, the suspension surfaces appear dry and rough and did not flow under tilt during or after sonication.  For $ 0.64 \ge \phi > 0.57$ they appear wetter but still rough.  In this range the samples did flow while being sonicated, but afterwards did not flow at infinitesimal tilt angles (measured at $\sim 1^{\circ}$).  However, for several of the packing fractions in this range there was a critical tilt angle $\Theta$ above which flow could be found.  For $\phi\le 0.57$, the samples appeared very shiny and smooth, and they flowed even at infinitesimal tilt angles.  

The critical packing fractions for the transitions in Fig.~\ref{fig:surfaceimages} match up with those usually found for the jamming transition in similar systems\citep{LN98}. The transition at $\phi=0.64$ corresponds to random close packing for frictionless spheres, above which packings of spheres are jammed and below which they can flow \citep{OSLN03}. When comparing to other published values for critical packing fractions, it is reasonable to expect an absolute uncertainty in the range of 0.01 in the packing fraction due to factors such as a finite size effect and polydispersity in the particle sizes, and sample preparation.  Experimentally it is usually found that packings of spheres remain mechanically stable, i.e. have a yield stress down to packing fractions as low as 0.56, called random loose packing, with the value of the packing fraction depending on friction and the density difference between the particles and fluid  \citep{OL90, JSSSSA08}.  Packing fractions closer to $\phi=0.64$ can only be reached for frictional, settling particles if the packings are vibrated which mobilizes the particle contacts, effectively eliminating the effect of friction \citep{KFLJN95}.  This explains why in the range  $0.64 \ge \phi > 0.57$ our suspensions flow under sonication but are otherwise jammed.  
This easily visible indicator of jamming based on surface roughness contrasts with other jammed systems, where there is no such visual test.  The insterstitial liquid allows this sensitive measurement by eye because changes in the fluid level at the surface due to changes in packing density only have to be on the scale of a particle size to dramatically change the surface appearance.  

We note the visible transition at $\phi=0.64$ in Fig.~\ref{fig:surfaceimages} is not always observed, depending on sample preparation.  If we do not sonicate, the samples can appear to be dry until they are diluted all the way down to $\phi=0.57$.  This could be because without the mobilization of particle contacts allowed by sonication, the suspension may not be able to pack more densely than $\phi=0.57$, perhaps trapping air bubbles, or to make up the extra space the liquid may retreat to the interior of the sample.    Despite this hysteresis effect in the visible transition at $\phi=0.64$ during preparation, rheology measurements with preshear consistently show the onset of a yield stress at the same packing fraction.

\begin{figure}                                                
\centerline{\includegraphics[width=2.75in]{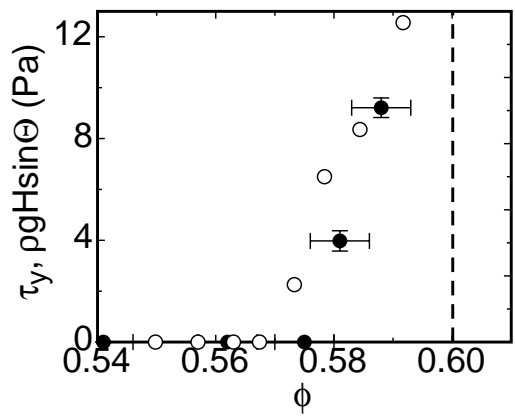}}
\caption{Open symbols:  gravitational stress $\rho g H\sin\Theta$ required to initiate shear under tilt for the suspensions shown in Fig.~\ref{fig:surfaceimages}.  The minimum tilt angle $\Theta$ required for shear is measured relative to a horizontal surface.  Dashed line:  threshold packing fraction above which the sample did not flow at any tilt angle (without sonication).  Solid symbols: yield stress $\tau_y$ obtained from viscosity curves in the rheometer.  The jamming transition at $\phi=0.57$ above which there is a yield stress coincides with the visible change in the surface shown in Fig.~\ref{fig:surfaceimages}.}
\label{fig:yieldstressangle}                                        
\end{figure}

We can connect the value of the critical tilt angle $\Theta$ for flow to the yield stress of the suspension.  For a tilt angle $\Theta$, the stress applied by gravity parallel to the surface is $\rho g H\sin\Theta$, which is shown in Fig.~\ref{fig:yieldstressangle} for packing fractions in which one was measured.  We compare this to the yield stress $\tau_y$ obtained from rheometer measurements in the zero shear rate limit.  For $\phi < 0.58$, we measured no yield stress, within our experimental resolution of $10^{-3}$ Pa.  There is an absolute uncertainty of about 0.005 in the packing fraction measurements  for the yield stress due to the process of loading the sample.  Within this uncertainty, the shear stress provided by gravity at the onset of shear under tilt matched with the measured yield stress in the rheometer.  This confirms that the visible changes in the surface can be quantitatively matched to  changes in the yield stress.  

The above analysis implies the macroscopic roughness of the surface of a suspension can be used as an indicator of the yield stress. At each packing fraction shown in Fig.~\ref{fig:surfaceimages} for $\phi > 0.57$, the sample surface has roughness on a macroscopic scale, i.e.~much larger than individual particles.  If an asperity of height $H$ forms on the upper surface of a fluid it can remain stable if the yield stress exceeds a value on the scale of $\rho g H$.  For these samples we observed asperities on the order of 1 mm at the highest packing fractions, consistent with the measured yield stresses on the order of 10 Pa and the value $H$ used in the gravitational stress scale.

The indication of a yield stress by the rough appearance of the surface confirms that surface tension forces are confining these suspensions at the boundary, whether or not there is trapped air.  Since the scale of forces from surface tension are similar at the boundary or due to trapped air, in general  it could be possible that the deformation at the liquid-air interface at the boundary is responsible for the yield stress, which is then transmitted along solid particle contacts through the suspension.  For our measurements, we know there is also trapped air based on volume measurements so the deformation of the liquid-air interface at the boundary is also balancing the stresses from surface tension in the interior.


\section{Normal stresses}
\label{sec:normalstress}

\begin{figure}                                                
\centerline{\includegraphics[width=3.4in]{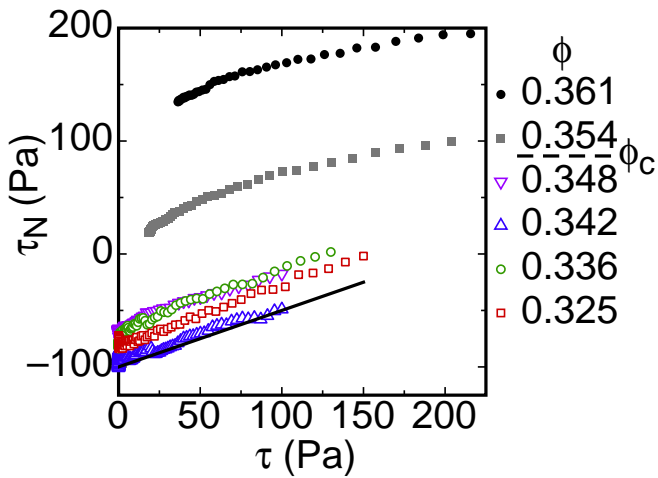}}
\caption{(color online) Normal stress $\tau_N$ vs.~shear stress $\tau$ for $\Gamma=9$ rods.  Packing fractions $\phi$ are shown in the key.  Solid line: linear relationship corresponding to a frictional response.  The dashed line in the key indicates the critical packing fraction $\phi_c$ above which the suspension is jammed with a large yield stress, and below which the suspension exhibits shear thickening (see Fig.~\ref{fig:viscstress_Gamma12}).
}
\label{fig:normalstress}                                        
\end{figure}

In this section we show the relation between the normal stresses and shear stresses.  We measured normal stresses along with the viscosity curves of Figs.~\ref{fig:viscstress_Gamma12}, \ref{fig:viscstress_Gamma6},  \ref{fig:viscstress_Gamma1}, and \ref{fig:viscstress_hookedrods}.  Some examples of normal stresses are plotted vs.~shear stresses in Fig.~\ref{fig:normalstress} for the $\Gamma=9$ rods at different packing fractions.  For packing fractions below $\phi_c$, the normal stress at zero shear stress is negative (pulling on the top plate), while above $\phi_c$, the normal stress is everywhere positive (pushing on the top plate).  For each packing fraction, we see  an increase in normal stress with increasing shear stress.  The roughly linear slopes and the fact that the shear and normal stress are on the same order of magnitude suggests a frictional scaling.  

\begin{figure}                                                
\centerline{\includegraphics[width=3.4in]{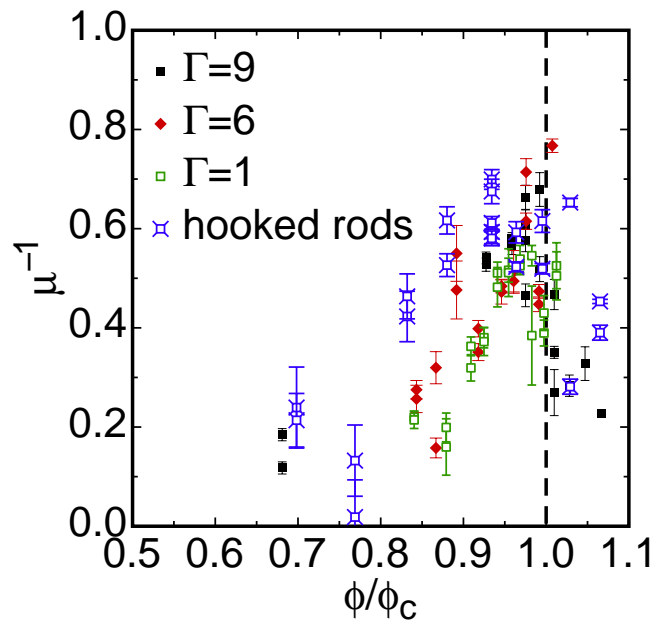}}
\caption{(color online) Inverse friction coefficient $\mu^{-1}$ from a fit of $\tau_N = \tau_{N,0} + \mu^{-1}\tau$.   Data are shown for several particle shapes listed in the key vs.~normalized packing fraction.  Significant values of $\mu^{-1}$, indicating a frictional relation, are observed for $\phi\stackrel{>}{_\sim}0.8\phi_c$ which is the same range where Discontinuous Shear Thickening is seen in Fig.~\ref{fig:epsilon_phirel}.}
\label{fig:friction}                                        
\end{figure}

\begin{figure}                                                
\centerline{\includegraphics[width=3.4in]{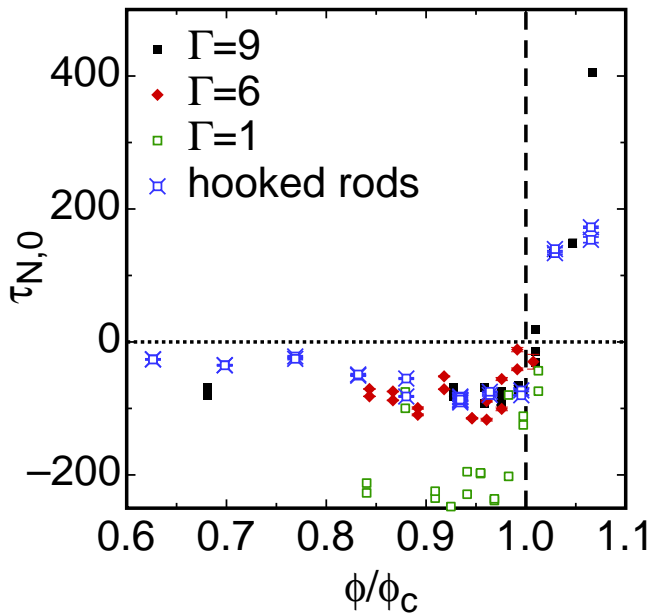}}
\caption{(color online) Zero-shear rate limit of the normal stress $\tau_{N,0} $ from a fit of $\tau_N = \tau_{N,0} + \mu^{-1}\tau$.  Data are shown for several particle shapes listed in the key vs.~normalized packing fraction.  Negative stresses for $\phi<\phi_c$ are expected due to surface tension at the edge of the suspension pulling the rheometer plates together, while positive stresses for $\phi>\phi_c$ are expected due to jamming of the suspension against the plates.}
\label{fig:nstressstatic}                                        
\end{figure}

To quantify such a frictional scaling, we use a fitting function $\tau_N = \tau_{N,0} + \mu^{-1}\tau$ with fit parameters $\mu$ corresponding to the effective dynamic friction coefficient and $\tau_{N,0}$ corresponding to the normal stress in the limit of zero shear stress.  This function is fit to the stress measurements for each viscosity curve shown in Figs.~\ref{fig:viscstress_Gamma12}, \ref{fig:viscstress_Gamma6},  \ref{fig:viscstress_Gamma1}, and \ref{fig:viscstress_hookedrods}.  The values of $\mu^{-1}$ and $\tau_{N,0}$ are plotted in Figs.~\ref{fig:friction} \ref{fig:nstressstatic}, respectively.  The error bars indicate the relative rms deviation of the linear fit from the data $\langle (\tau_{N,0} + \mu^{-1}\tau-\tau_N)^2\rangle^{1/2}/[max(\tau)-min(\tau)]$, so any strongly non-linear relationship would be revealed by large error bars.

The fit values of $\mu^{-1}$ in Fig.~\ref{fig:friction} tend to be small at low packing fractions, then increase and level off at around 0.5 as $\phi$ approaches $\phi_c$ from below, then suddenly drop again below $\phi_c$.  The linear relationship between shear and normal stress, with values of the friction coefficient $\mu$ on the order of 1 are typical of a frictional relationship, commonly seen in Discontinuous Shear Thickenining \cite{LDH03, LDHH05, BJ10}.  It is notable that we never measure a significantly negative $\mu$ in Discontinuous Shear Thickening, which might be expected from purely hydrodynamic models for shear thickening \cite{BBV02}.  The variation of $\mu$ with $\phi/\phi_c$ suggests a couple of different transitions.  In the Newtonian limit at low packing fractions, $\mu^{-1}$ is expected to go to zero as the shear flow should not produce any normal force in a Newtonian fluid.  The increase in $\mu^{-1}$ with $\phi/\phi_c$ suggests an increasing non-Newtonian contribution to the stresses up to a dense limit.  A similar transition to positive normal stresses which are linear in the shear stress at high packing fractions has been seen previously for spherical particles \cite{LDHH05}.  This transition where frictional forces are significant corresponds to the same packing fraction range of $\phi\stackrel{>}{_\sim}0.8\phi_c$ where Discontinuous Shear Thickening tends to be become dominant over other viscous or inertial forces as in Fig.~\ref{fig:epsilon_phirel} or \cite{BBV02}.  This correspondence supports the idea that frictional forces are responsible for Discontinuous Shear Thickening \cite{BJ10}.  At lower $\phi$, it is presumably more difficult to produce system-spanning chains of solid contacts between particles.

The fit values of the static limit of the normal stress $\tau_{N,0}$ in Fig.~\ref{fig:nstressstatic} are all negative for $\phi < \phi_c$.  Observation of the suspensions when loaded into a parallel plate rheometer reveals a concave curvature of the liquid-air interface at the side, which produces stress from surface tension to pull the rheometer plates together.  The scale of this stress is expected to be about $-\gamma/d \approx -50$ N, on the same order as seen in Fig.~\ref{fig:nstressstatic}.  On the other hand, increasingly positive normal stresses are seen in the static limit for $\phi > \phi_c$.  Above the jamming  transition, particles are expected to push back against each other as they are confined to a tight packing \cite{LN98}.  As a result, they are also expected to push against the top plate, resulting in the positive normal forces observed.  The observations of this transition to positive normal stresses and the change in friction coefficient $\mu$ at the same packing fraction where the yield stress jumps (our definition of $\phi_c$) are consistent with expectations that both shear and normal stress scalings change sharply at the jamming transition \cite{OLLN02}.

\section{Discussion}
\label{sec:discussion}

Here we comment on some earlier works which we may be able to understand better based on our more detailed studies of particle shape effects on the strength of Discontinuous Shear Thickening.  

Clarke measured viscosity curves for several different shapes at a fixed packing fraction \cite{Clarke67} .  He reported the order of steepest to shallowest slope of viscosity curves: rods, plates, grains, spheres.  Based on this he concluded that more anisotropic shapes shear thicken more strongly.  
In comparison, we showed in Fig.~\ref{fig:epsilon_phirel} that the steepness of viscosity curves of different convex shapes collapse when the packing fraction is normalized by the critical packing fraction $\phi_c$ such that the steepness becomes greater approaching $\phi_c$ from below.  This implies that when different shapes are compared at the same packing fraction, the ones with the smallest $\phi_c$ will be closer to $\phi_c$ and therefore steeper.  Based on these results it seems more complete to identify the effect of particle shape as changing the value of the critical point $\phi_c$, where more anisotropic shapes tend to produce lower values of $\phi_c$.  


Suspensions have been found to exhibit weaker shear thickening with increasing particle size polydispersity at a fixed packing fraction \cite{KK91}.  In contrast, we can collapse data in Fig.~\ref{fig:epsilon_phirel} as a function of normalized packing fraction without regard to polydispersity, which ranged from very uniform for the PRINT$^{\textregistered}$ process, to very polydisperse cornstarch suspensions.  We suggest that the effect observed by Ref.~\cite{KK91} was a shift in the value of $\phi_c$, which is known to shift upwards for more polydisperse packings \cite{OSLN03}.  In fact, Ref.~\cite{KK91} reported the value of $\phi_c$ based on sediment concentrations, and performed measurements over a range of packing fractions, but still measured much closer to $\phi_c$ for the low polydispersity suspensions than for the high polydispersity suspensions.  Thus, the weakening observed by Ref.~\cite{KK91} for increasing polydispersity at fixed packing fraction can be explained as a result of moving further away from $\phi_c$.

\section{Conclusions}
\label{sec:conclusions}

We measured the rheology of dense suspensions of fabricated PEG particles with different shapes, and found Discontinuous Shear Thickening that is qualitatively similar for each particle shape (Figs.~\ref{fig:viscstress_Gamma12}, \ref{fig:viscstress_Gamma6},  \ref{fig:viscstress_Gamma1}, and \ref{fig:viscstress_hookedrods}).  The logarithmic slopes of the stress/shear-rate relation in the shear thickening regime for convex particles are found to collapse onto a single curve when the packing fraction $\phi$ is normalized by $\phi_c$, which corresponds to the jamming transition measured independently as the onset of a large yield stress (Fig.~\ref{fig:epsilon_phirel}).  Additionally, these slopes diverge in the limit as $\phi$ approaches $\phi_c$ from below, as was found for nearly spherical particles \cite{BJ09}, suggesting the jamming transition controls Discontinuous Shear Thickening like a critical point.

The fact that $\phi/\phi_c$ acts as the control parameter for the strength of Discontinuous Shear Thickening has several consequences.  It suggests that the apparent weakening of shear thickening for more spherical shapes \cite{Clarke67} or more polydisperse sizes \cite{KK91} at constant $\phi$ can be explained by the fact that $\phi_c$ is moved further away.  Similarly, particles that form fractal aggregates have been found to exhibit Discontinuous Shear Thickening at packing fractions as low as 11\%, which is not surprising in this interpretation since they also pack at very low densities \cite{OMY10}.  For designer suspensions, this implies that strong shear thickening can be achieved using less material by designing particles that pack at low densities.

For one shape we studied, the hooked rods, the slopes of the viscosity curves did not collapse with the same scaling as for convex particles.  We can only speculate on why hooked rods show a different scaling.  The non-convex shape allows for more contact points, which can frustrate motions in both shear and compressional directions on contact, rather than just compression for convex particles.  The hooks may frustrate particle rotation and passing in tight packings under shear flow, but it is not clear how these features specifically translate to the wider packing fraction range for strong shear thickening.

We identified two other mechanical properties which relate Discontinuous Shear Thickening to jammed systems.  First, the shear stress and normal stress are linearly related as in a frictional scaling for $\phi\stackrel{>}{_\sim}0.8\phi_c$ where Discontinuous Shear Thickening and jamming occur (Figs.~\ref{fig:normalstress}, \ref{fig:friction}), which suggests that energy dissipation comes from friction between solid particle contacts that span the system in both cases, rather than from viscous dissipation.  Additionally, we showed the scale of the yield stress comes from interfacial tension for jammed suspensions, which can be seen directly by observing the surface roughness of the suspension (Fig.~\ref{fig:yieldstressair}, \ref{fig:surfaceimages}, \ref{fig:yieldstressangle}).  This is similar to when the suspensions dilate against the liquid-air interface to result in Discontinuous Shear Thickening, although generally the forces need not come from interfacial tension, but could come from any forces which confine the system at the boundary \cite{BJ10}.  The difference between the jammed state above $\phi_c$ and the Discontinuous Shear Thickening regime is that the latter requires dilation from shear to achieve the solid particle contacts to reach the frictional state, while the jammed state already has solid particle contacts at rest.  The similarity in the mechanics of force transmission suggests the possibility that a generalization of jamming theory may also be able to describe Discontinuous Shear Thickening systems, as was suggested by \cite{CWBC98}.

\section{Acknowledgements}

Thanks to Carlos Orellana for performing preliminary rheology measurements on PEG particles to help us settle on a formulation.  This work was supported by DARPA through Army grant W911NF-08-1-0209 and by the NSF MRSEC program under DMR-0820054.


\begin{thebibliography}{}

\bibitem{MW58} A. B. Metzner and M. Whitlock, Trans. Soc. Rheol. {\bf 11}, 239 (1958). 

\bibitem{Ho72} R. L. Hoffmann, Trans. Soc. Rheol. {\bf 16}, 155 (1972).

\bibitem{Ba89} H. A. Barnes, J. Rheology {\bf 33}, (2) 329 (1989).

\bibitem{MW01a} B. J. Maranzano and N. J. Wagner,  J. Chem. Phys. {\bf 114}, (23) 10514 (2001).

\bibitem{BJ10}E. Brown and H. M. Jaeger,  submitted to J. Rheology, http://arxiv.org/abs/1010.4921.

\bibitem{LWW03} Y. S. Lee, E. D. Wetzel, and N. J Wagner, J. Materials Sci. {\bf 38}, 2825 (2003).


\bibitem{Philipse96} A. P. Philipse, Langmuir {\bf 12}, 1127 (1996).

\bibitem{EW05} R. G. Egres and N. J. Wagner,  J. Rheol. {\bf 49} (3) 719 (2005).

\bibitem{CWBC98} M. E. Cates, J.P. Wittmer, J.-P. Bouchaud, and P. Claudin, Phys. Rev. Lett. {\bf 81} (9), 1841 (1998). 

\bibitem{BJ09} E. Brown, and H. M. Jaeger, Phys. Rev. Lett. {\bf 103}, 086001 (2009).

\bibitem{LN98}A. J. Liu and S. R. Nagel,  Nature {\bf 396}, 21 (1998).



\bibitem{Clarke67} B. Clark, Trans. Inst. Chem. Eng. {\bf 45}, 251 (1967).

\bibitem{Beazley80} K. M. Beazley, K. M., Rheometry: Industrial Applications, edited by K. Walters 
(Research Studies Press, Chichester, 1980), 339.

\bibitem{RMEEDD05} J. P. Rolland, B. W. Maynor, L. E. Euliss, A. E. Exner, G. M. Denison, J. M. DeSimone,  J. Am. Chem. Soc. {\bf 127}, (28) 10096 (2005).

\bibitem{BFOZMBDJ10} E. Brown, N. A. Forman, C. S. Orellana, H. Zhang, B. W. Maynor, D. E. Betts, J. M. DeSimone, and H. M. Jaeger, Nature Materials {\bf 9}, (3) 220 (2010).

\bibitem{BZFMBDJ10} E. Brown, H. Zhang, N. A. Forman, B. W. Maynor, D. E. Betts, J. M. DeSimone, and H. M. Jaeger,  J. Rheology {\bf 54}, 1023 (2010).



\bibitem{OT07} P. Olsson and S. Teitel, Phys. Rev. Lett. {\bf 99}, 178001 (2007).


\bibitem{OLLN02} C. S. O'Hern, S. A. Langer, A. J. Liu, and S. R. Nagel, Phys. Rev. Lett. {\bf 88}(7), 075507 (2002).

\bibitem{TPCSW01} V. Trappe, V. Prasad, L. Cipelletti, P. N. Segre, and D. A. Weitz, Nature {\bf 411} 772 (2001).

\bibitem{GDJC98}  B.S. Gardiner, B. Z. Dlugogorski, G. J. Jameson, and R. P. Chhabra, J. Rheol {\bf 42} (6), 1437 (1998).

\bibitem{KW11} E. Koos and N. Willenbacher, Science {\bf 331} (6019), 897 (2011).


\bibitem{OSLN03} C. S. O'Hern,  L. E. Silbert, A. J. Liu, and S. R. Nagel, Phys. Rev. E {\bf 68} (1), 011306 (2003).

\bibitem{KFLJN95} J.B. Knight,, C.G. Fandrich, C.N. Lau, H.M. Jaeger, and S.R. Nagel, ``Density relaxation in a vibrated granular material," Phys. Rev. E {\bf 51} (5), 3957 (1995).

\bibitem{CHH05} M.E. Cates, M.D. Haw and C.B. Holmes, J. Phys: Condens. Matter {\bf 17} S2517 (2005).




\bibitem{OL90} G.Y. Onoda and E. G. Liniger, Phys. Rev. Lett. {\bf 64} (22), 2727 (1990).

\bibitem{JSSSSA08} M. Jerkins, M. Schr{\"o}ter, H.L. Swinney, T.J. Senden, M. Saadatfar, and T. Aste, Phys. Rev. Lett. {\bf 101}, 018301 (2008).

\bibitem{LDH03}  D. Lootens, H. Van Damme, and P. H{\'e}braud, Phys. Rev. Lett. {\bf 90}, No. 17, 178301 (2003). 

\bibitem{LDHH05} D. Lootens, H. Van Damme, Y. H{\'e}mar, and P. H{\'e}braud,  Phys. Rev. Lett. {\bf 95}, 268302 (2005).

\bibitem{BBV02} J. Bergenholtz, J. F. Brady, and M. Vivic,  J. Fluid Mech. {\bf 456}, 239 (2002). 


\bibitem{KK91} D. S. Keller,  and D.V. Keller,  J. Rheol. {\bf 35}(8), 1583 (1991).


\bibitem{OMY10} B. Ozel, Y. Z. Menceloglu, and M. Yildiz, submitted to Langmuir (2010).







\end{thebibliography}
\end{document}